\documentclass[11pt, a4paper]{article}

\usepackage{jheppub}
\usepackage{amsmath}
\usepackage{amsfonts}
\usepackage{amssymb} 
\usepackage{mathrsfs}
\usepackage[utf8]{inputenc}
\usepackage{hyperref}

\title{Heterotic String Field Theory with Manifest Spacetime Supersymmetry}
\author{Nathan Berkovits, Ulisses M. Portugal}

\affiliation{ICTP South American Institute for Fundamental Research\\Instituto de F\'{i}sica Te\'{o}rica, UNESP — Universidade Estadual Paulista,\\Rua Dr. Bento T. Ferraz 271, 01140-070, S\~{a}o Paulo, SP, Brasil}
\emailAdd{nathan.berkovits@unesp.br}
\emailAdd{ulisses.portugal@unesp.br}

\abstract{ Using the hybrid formalism with manifest $N=1$ $d=4$ spacetime supersymmetry, we construct the quadratic term in the heterotic superstring field theory action. As in open superstring field theory using the hybrid formalism, the heterotic string field theory action is constructed with three string fields and the massless sector describes $N=1$ $d=10$ supergravity in terms of $N=1$ $d=4$ superfields.}

\begin{document}

\maketitle

\section{Introduction}
A classical action for the Neveu-Schwarz sector of heterotic superstring field theory (SFT) was first constructed in \cite{Berkovits:2004xh,Okawa:2004ii}. The basic ingredients are the WZW-like large Hilbert space formulation of open superstring field theory of \cite{Berkovits:1995ab} and the closed string field products \cite{Zwiebach:1992ie}. The heterotic field can be thought of as a left-right product of an open superstring field and an open bosonic string field. This construction was extended to include the Ramond sector up to quartic order in fermions in \cite{Goto:2016ckh}.

There is a formulation of open superstring field theory using the hybrid formalism \cite{Berkovits:1995ab} which has manifest four dimensional super-Poincaré invariance. This formulation is useful for describing Calabi-Yau compactifications of the superstring, and provides a SFT generalization of the superspace formulation of $d=10$ super Yang-Mills in terms of $N=1$ $d=4$ superfields \cite{Marcus:1983wb}. It would be useful to have a similar formulation of heterotic closed SFT, which at the massless level would correspond to a superspace formulation of $N=1$ $d=10$ supergravity in terms of four-dimensional superfields. In this paper we take a first step towards this goal and construct the quadratic term in the heterotic string field theory action with manifest $N=1$ $d=4$ super-Poincaré invariance in terms of three string fields, similar to what was done for the open superstring. We then analyze explicitly the massless contribution to the action and show that, upon restricting to the Calabi-Yau (CY) independent sector, it provides a description of four dimensional supergravity plus a tensor multiplet in $N=1$ $d=4$ superspace. Work is in progress on extending this quadratic action to the full non-linear closed string field theory action of the heterotic superstring.

In section \ref{sec Linearized action} we obtain the linearized equations of motion and gauge invariances describing the heterotic spectrum, and construct the corresponding quadratic action. In section \ref{sec Massless level} we explicitly evaluate the action for the massless sector and show that the CY independent sector correctly describes supergravity in $N=1$ $d=4$ superspace. We also compare with the RNS formulation of heterotic string field theory. Finally, we discuss our results in section \ref{sec Discussion}.

\section{Linearized action}\label{sec Linearized action}
In this section we will find that all the physical states of the heterotic string can be described using three string fields, and will construct a quadratic action for heterotic string field theory with manifest four dimensional spacetime supersymmetry.

We take the string field $\Phi$ to be a direct product of a bosonic open string in the antiholomorphic sector and a hybrid open superstring in the holomorphic sector. $\Phi$ is grassmann odd with total ghost number $1$, i.e. $\oint(-\partial\rho+J_C-\overline b\overline c)\Phi = \Phi$ where $\rho$ is the chiral boson in the hybrid formalism and $J_C =-i \partial H_C$ is the CY charge. Following the notation of \cite{Berkovits:1995ab}, on-shell states should satisfy the linearized equation of motion 
\begin{equation}
    \tilde{G}^{+} (G^++\overline Q) \Phi=0
\end{equation}
where conformal weight one operators act by contour integration around the string field, $\overline Q= \int d\overline z (\overline c \overline T+ \overline c \overline\partial \overline c \overline b)$ is the antiholomorphic BRST operator, $G^+ = G_4^+ + G_6^+$ and $\tilde{G}^+ = \tilde {G}_4^+ + \tilde{G}_6^+$ are the holomorphic BRST charge and $\eta$ ghost in the RNS formalism and, after performing a field redefinition to the $N=1$ $d=4$ hybrid formalism, 

\begin{equation}\label{subsidiarya}
    G_4^+ = d^2 e^\rho, \quad G_6^+ = \psi^j \partial x_j, \quad
\tilde {G}_4^+ = \overline d^2 e^{-2\rho + i H_C}, \quad \tilde {G}_6^+ = \frac{1}{2} e^{-\rho}\epsilon_{jkl} \partial x^j\psi^k \psi^l.
\end{equation}
In addition, the string field should satisfy the hybrid version of the $b_0^-$ and $L_0^-$ conditions on closed string fields which are
\begin{equation}\label{subsidiary}
    (G^- - \overline b)_0 \Phi = 0, \quad L_0^-\Phi = 0
\end{equation}
where $G^- = G_4^- + G_6^-$ and
\begin{equation}\label{subsidiaryb}
 G_4^- =  \overline d^2 e^{-\rho}, \quad G_6^-=  \overline\psi_j \partial x^j.
\end{equation}
Here we run into an issue with the reality condition: in the hybrid formalism, hermitian conjugation (in Minkowski signature) is defined in such a way that $G^{\pm\dagger}=\tilde G^{\pm}$, and then the first condition of \ref{subsidiary} is evidently not real. For now we will simply work in signature $d=(2,2)$ or $d=(5,5)$ so that all the operators and fields are real. It will later be shown that the physical spectrum can be analytically continued to Minkowski signature.

To fix the ambiguity that comes from picture changing, rather than fixing the picture number, we consider the more restrictive equation
\begin{equation}\label{eom}
    (G^+ +\overline Q+\tilde{G}^{+}) {\Phi}=0
\end{equation}
with the gauge invariance
\begin{equation}
    \delta {\Phi}=(G^+ +\overline Q+\tilde{G}^{+}) {\Lambda}
\end{equation}
just as was done for the open superstring. The gauge parameter $\Lambda$ should satisfy the same conditions \ref{subsidiary}.

Now the idea would be to expand $\Phi$ in eigenvalues of the $\rho$ charge:
\begin{equation}
    {\Phi}=\sum_{n=-\infty}^{\infty} \Phi_{n}
\end{equation}
where $-\partial\rho\Phi_n = n\Phi_n$. Our hope is that we can describe $\Phi$ in terms of only a finite number of the $\Phi_n$'s. But here we encounter a problem in the fact that $G^- - \overline b$ does not have a well defined $\rho$-charge, which means that the subsidiary condition will mix different $\Phi_n$. Explicitly, we have $(G_4^-)_0\Phi_n = - (G_6^--\overline b)_0\Phi_{n-1}$.

To avoid this, we will solve the $(G^--\overline b)_0$ constraint as
\begin{equation}
    \Phi = (G^- - \overline b)_0\Sigma
\end{equation}
and work with $\Sigma$ instead of $\Phi$. Note that we can choose $\Sigma$ to be annihilated by $\overline c_0$, and than it is expressed in terms of $\Phi$ as $\Sigma = \overline c_0\Phi$. Now we have a constraint $\overline c_0$ with a well defined $\rho$-charge. The price we pay is that the equations for $\Sigma$ take a more complicated form, namely \ref{eom} implies that
\begin{equation}
    (G^+ + \overline Q + \tilde G^+ + 2\sum_{n=0}^\infty n \overline c_{-n}\overline c_n (G^- - \bar b)_0)\Sigma = 0
\end{equation}
with gauge invariance
\begin{equation}
    \delta\Sigma = (G^+ + \overline Q + \tilde G^+ + 2\sum_{n=0}^\infty n \overline c_{-n}\overline c_n (G^- - \bar b)_0)\Lambda
\end{equation}
where $\Lambda$ is also annihilated by $\overline c_0$. Note that $(G^+ + \overline Q + \tilde G^+ + 2\sum_{n=0}^\infty n \overline c_{-n}\overline c_n (G^- - \bar b)_0)$ is nilpotent (in the space of string fields annihilated by $L_0^-$) and anticommutes with $\overline c_0$. Now we can expand $\Sigma$ in eigenvectors of the $\rho$-charge, $\Sigma = \sum_{-\infty}^\infty\Sigma_n$, with the simple constraints
\begin{equation}
    \overline c_0 \Sigma_n = 0, \quad L_0^-\Sigma_n = 0
\end{equation}

The equations of motion and gauge invariances become
\begin{equation}\label{sigma eom}
    G'^+_4 \Sigma_{n} + G'^{+}_{6} \Sigma_{n+1} + \tilde G'^{+}_{6} \Sigma_{n+2}+\tilde G'^{+}_{4} \Sigma_{n+3}=0
\end{equation}
\begin{equation}
    \delta \Sigma_{n}=G'^+_4 \Lambda_{n-1}+G'^{+}_{6} \Lambda_{n}+\tilde G'^{+}_{6} \Lambda_{n+1}+\tilde G'^{+}_{4} \Lambda_{n+2}
\end{equation}
where
\begin{gather}
    G'^+_4 = G_4^+, \quad \tilde G'^{+}_{4} = \tilde G_4^+, \\
    G'^{+}_{6} = G_6^+ + \overline Q +  2\sum_{n=0}^\infty n \overline c_{-n}\overline c_n (G^-_6 - \overline b)_0, \quad \tilde G'^{+}_{6} = \tilde G_6^+ +  2\sum_{n=0}^\infty n \overline c_{-n}\overline c_n (G^-_4)_0
\end{gather}
Note that each of $G'^+_4$, $G'^+_6$, $\tilde G'^+_6$ and $\tilde G'^+_4$ anti-commutes with $\overline c_0$. Note also that $G'^+_4$ and $\tilde G'^+_4$ are nilpotent and have trivial cohomology, which means we can use \ref{sigma eom} to write all $\Sigma_n$ in terms of only three, which we choose to be $\Sigma_{-1}$, $\Sigma_{0}$ and $\Sigma_{1}$. 

The equations of motion for these three string fields are
\begin{gather}\label{reduced eom}
    \tilde G'^+_4(G'^+_4\Sigma_{-1} + G'^+_6 \Sigma_0 + \tilde G'^+_6\Sigma_1) = 0 \\
    (\tilde G'^+_6 G'^+_6 +\tilde G'^+_4 G'^+_4)\Sigma_0 - G'^+_6 \tilde G'^+_4 \Sigma_1 + \tilde G'^+_6 G'^+_4\Sigma_{-1} = 0 \label{reduced eom 2} \\
    G'^+_4(G'^+_6\Sigma_{-1} + \tilde G'^+_6\Sigma_0+\tilde G_4^+\Sigma_1) = 0 \label{reduced eom 3}
\end{gather}
with linear gauge transformations
\begin{gather}\label{gauge linear}
    \delta\Sigma_{-1} =  G'^+_4\Lambda_{-2} + G'^+_6\Lambda_{-1} + \tilde G'^+_6\Lambda_0 + \tilde G'^+_4 \Lambda_1 \\
    \delta \Sigma_0 = G'^+_4\Lambda_{-1}+G'^+_6\Lambda_0+\tilde G'^+_6\Lambda_1+\tilde G'^+_4\Lambda_2 \\
    \delta\Sigma_1 = G'^+_4\Lambda_0 + G'^+_6\Lambda_1 + \tilde G'^+_6\Lambda_2 + \tilde G'^+_4\Lambda_3 \label{gauge linear 3}
\end{gather}
with
\begin{equation}
    \overline c_0\Lambda_n = 0, \quad L_0^-\Lambda_n = 0
\end{equation}
We should note that $G'^+_6$ and $\tilde G'^+_6$ are not nilpotent like $G^+_6$ and $\tilde G^+_6$, and also that $G'^+_4$ does not anti-commute with $\tilde G'^+_6$ and $G'^+_6$ with $\tilde G'^+_4$. Explicitly, we have
\begin{gather}
    (G'^+_6)^2 = 2\sum_{n=0}^\infty n\overline c_{-n}\overline c_n (L_6 - \overline L)_0, \\
    (\tilde G'^+_6)^2 = -\{\tilde G'^+_4,G'^+_6\} = -2\sum_{n=0}^\infty n\overline c_{-n}\overline c_n e^{-2\rho}\epsilon_{ijk}\psi^i\psi^j\partial x^k\overline d^2, \\
    \{G'^+_4,\tilde G'^+_6\} = 2\sum_{n=0}^\infty n\overline c_{-n}\overline c_n (L_4)_0, \\
    \{G'^+_4,\tilde G'^+_4\} = -\{G'^+_6,\tilde G'^+_6\} = \partial\rho e^{-\rho+iH_C}
\end{gather}
where $L_4$ and $L_6$ are the four-dimensional and six-dimensional parts of the holomorphic energy-momentum tensor, respectively, and $\overline L$ is the antiholomorphic energy-momentum tensor. All other anticommutators vanish. Still, the fact that $G'^+_4+G'^+_6+\tilde G'^+_6+\tilde G'^+_4$ is nilpotent ensures gauge invariance.

Now we would like to define a quadratic action whose variation gives the above equations. In order to do this, we define an inner product with a $\overline b_0$ insertion, $\langle A,B\rangle \equiv \langle A|\overline b_0 |B\rangle$. Although $G'^+_6$ does not anti-commute with $\overline b_0$, the $\overline c_0$ constraint ensures that the anti-commutator vanishes inside the brackets, such that
\begin{equation}
    \langle A, G' B\rangle = (-1)^A \langle G'A, B\rangle
\end{equation}
where $G'$ stands for any of $G'^+_4$, $G'^+_6$, $\tilde G'^+_6$ and $\tilde G'^+_4$, and $(-1)^A$ denotes the grassmanality of $A$.

With this definition, the action which reproduces the equations \ref{reduced eom}-\ref{reduced eom 3} and gauge transformations of \ref{gauge linear}-\ref{gauge linear 3} is
\begin{gather}\label{quadratic action}
    S_2 = \frac{1}{2}\left(\left<\tilde G'^+_4\Sigma_0,G'^+_4\Sigma_0\right> + \left<\tilde G'^+_6\Sigma_0,G'^+_6\Sigma_0\right> + \left<\Omega,\tilde G'^+_6\Sigma_1\right> + \left<G'^+_6\Sigma_{-1},\overline\Omega\right>\right) \\
    + \left<\tilde G'^+_6\Sigma_0,\overline\Omega\right> + \left<\Omega,G'^+_6\Sigma_0\right> + \left<\Omega,\overline\Omega\right>\nonumber
\end{gather}
where $\Omega=\tilde G'^+_4\Sigma_1$, $\overline\Omega=G'^+_4\Sigma_{-1}$. This is formally the same as the open superstring field action \cite{Berkovits:1995ab}.

Note that we can also define string fields of ghost number one which are annihilated by $\bar b_0$ as $\Phi'_n = \bar b_0\Sigma_n$. Then, if we redefine the inner product as $\langle A,B\rangle = \langle A|\bar c_0|B\rangle$, we can write the action \ref{quadratic action} in the same form in terms of $\Phi'_n$. This form is more similar to the usual formulations of closed string field theories, and in particular the RNS formulation of heterotic SFT, so it might be more suitable for constructing the nonlinear orders. Note however that $\bar b_0$ does not anti-commute with $G'^+_6$, so the gauge transformations do not take the same form - instead there will be a $\bar b_0$ in front of every term in \ref{gauge linear}, and the parameters $\Lambda_n$ are still annihilated by $\bar c_0$.

\section{Massless level}\label{sec Massless level}
Having constructed a quadratic action, it will be interesting to analyze it explicitly for the massless level. We expect this should give a description of ten-dimensional supergravity in terms of four-dimensional superfields. We will ignore the right-moving fermionic worldsheet variables, that is the states transforming under $E8\times E8$ or $SO(32)$. Then the most general form of the string fields at the massless level is
\begin{gather}
    \overline b_0\Sigma_0 = \overline c\overline\partial x_MH^M(x,\theta,\overline\theta) + \psi^j C_j(x,\theta,\overline\theta) \\
    \overline b_0\Sigma_1 = e^\rho F + e^\rho\overline\psi_j\overline c\overline\partial x^M\overline A^j_M(x,\theta,\overline\theta) \\
    \overline b_0\Sigma_{-1} = e^{-\rho}\overline c\overline\partial^2\overline cB(x,\theta,\overline\theta) + e^{-\rho}\psi^j \overline c\overline\partial x_M A_j^M(x,\theta,\overline\theta) + \frac{1}{2}e^{-\rho}\psi^i\psi^j\epsilon_{ijk}\overline C^k(x,\theta,\overline\theta)
\end{gather}
where $M$ runs over all the ten dimensions, while $i,j,k$ are $SU(3)$ indices describing the six compactified directions. All the fields in the above expansions have conformal weight zero, except for $F$ which has conformal weight $1$. The graviton is contained in the fields $H^M$, $A^M_j$ and $\overline A^j_M$, while $B$ is related to the ghost dilaton. For the gauge parameters, we have:
\begin{gather}
    \bar b_0\Lambda_{-2} = e^{-2\rho}\partial\theta^{\alpha}\left(\overline c\overline\partial^2\overline c \omega_{\alpha}(x,\theta,\overline\theta) + \psi^j\overline c\overline\partial x_M\omega^M_{j\alpha}(x,\theta,\overline\theta) + \frac{1}{2}\epsilon_{ijk}\psi^i\psi^j\overline\omega^k_{\alpha}(x,\theta,\overline\theta)\right) \\
    \bar b_0\Lambda_{-1} = e^{-\rho}\left(\overline c\overline\partial x_M\overline N^M(x,\theta,\overline\theta) + \psi^j\lambda_j(x,\theta,\overline\theta)\right) \\
    \bar b_0\Lambda_0 = P(x,\theta,\overline\theta) \\
    \bar b_0\Lambda_1 = e^\rho\overline\psi_j\overline\lambda^j(x,\theta,\overline\theta) \\
    \bar b_0\Lambda_2 = \frac{1}{2}e^{2\rho}\epsilon^{ijk}\overline\psi_i\overline\psi_j\partial\overline\theta^{\dot\alpha}\omega_{k\dot\alpha}(x,\theta,\overline\theta) + e^{2\rho-iH_C}\overline c\overline\partial x_MN^M(x,\theta,\overline\theta) \\
    \bar b_0\Lambda_3 = e^{3\rho-iH_C}\left(\Psi + \partial\overline\psi_j\partial\overline\theta^{\dot\alpha}\overline c\overline\partial x^M\omega^j_{M\dot\alpha}(x,\theta,\overline\theta)\right)
\end{gather}
Here we have used some of the gauge for gauge freedom to fix this form of the gauge parameters. All the parameters in the above expansions have conformal weight zero, except for $\Psi$ which has conformal weight $3$. This parameter can be used to fix the form of $F$ to
\begin{equation}
    F = \partial\overline\theta^{\dot\alpha}\overline\beta_{\dot\alpha}(x,\theta,\overline\theta)
\end{equation}
leaving the residual gauge symmetry
\begin{equation}
    \delta\overline\beta_{\dot\alpha} = \overline D^2\alpha_{\dot\alpha}
\end{equation}

Evaluating the action \ref{quadratic action} at the massless level, we get
\begin{gather}\label{hybrid massless action} 
    S_2 = \frac{1}{8}\int d^{10}xd^4\theta \\
    \left[ H^M \left(D\overline D^2 D H_M + 4\partial^j\partial_j H_M - 2\partial_M\overline D\overline\beta - 8\overline D^2\partial_k\overline A^k_M - 8D^2\partial^j A_{jM} - 8\partial_M\partial^jC_j\right) + \right.\nonumber\\
    C_i\left(8\partial^i\partial^jC_j + 2\sqrt{2}\epsilon^{ijk}\overline D^2\partial_jC_k + 8\partial^M\overline D^2\overline A_M^i + 4\partial^iD^2B - 2\overline D^2D^2\overline C^i + 4\partial^i\overline D\overline\beta\right) + \nonumber \\
    D^2\overline C^i\left(-4\partial_iB + 8\partial_M A^M_i - 2\sqrt{2}\epsilon_{ijk}\partial^j\overline C^k\right) - 8\sqrt{2}D^2  A_i^M\epsilon^{ijk}\partial_j 
    A_{kM} + \nonumber \\
    \left.\overline D^2 \overline A^i_M \left(-8D^2 A^M_i + 8\sqrt{2}\epsilon_{ijk}\partial^j\overline A^{kM}\right) - \overline D\overline\beta\left(D^2 B + \frac{1}{2}\overline D\overline\beta\right)\right] \nonumber
\end{gather}
which is invariant under the gauge transformations
\begin{gather}
    \delta H^M = D^2 \overline N^M + \overline D^2 N^M + \partial^M P \label{linear gauge 1}\\
    \delta A_i^M = \partial_i \overline N^M + \partial^M \lambda_i + D\omega^M_i \label{linear gauge 2}\\
    \delta \overline A^i_M = \partial^iN_M + \partial_M\overline\lambda^i + \overline D\overline\omega^i_M \label{linear gauge 3}\\
    \delta C_i = \partial_i P - D^2\lambda_i + 2\sqrt{2}\epsilon_{ijk}\partial^j\overline\lambda^k + \overline D\omega_{i} \label{linear gauge 4}\\
    \delta\overline C^i =  \partial^i P - 2\sqrt{2}\epsilon^{ijk}\partial_j\lambda_k - \overline D^2\overline\lambda^i + D\overline\omega^i \label{linear gauge 5}\\
    \delta B = 2\partial_M\overline N^M -\frac{1}{2}\overline D^2 P + 4\partial^j\lambda_j + D\omega \label{linear gauge 6}\\
    \delta\overline\beta_{\dot\alpha} = D^2\overline D_{\dot\alpha}P - 4\overline D_{\dot\alpha}\partial_j\overline\lambda^j - 4\partial^j\omega_{j\dot\alpha} + \overline D^2\alpha_{\dot\alpha}\label{linear gauge 7}
\end{gather}

As mentioned above, we expect this to describe ten-dimensional supergravity. This is not immediately clear, especially given that the action \ref{hybrid massless action} is not even real in Minkowski signature. To simplify the problem, we look in the next subsection at the compactification independent part of the action.

\subsection{Calabi-Yau independent states}
We now consider only the CY independent states, by which we mean the massless string fields that do not depend on $\psi^j$, $\overline{\psi}_j$, $x^j$ or $\overline{x}_j$. That leaves us with
\begin{equation}
    \overline b_0\Sigma_0 = H^m(x,\theta,\overline\theta)\overline c\overline\partial x_m, \quad \overline b_0\Sigma_{-1} = e^\rho F, \quad \overline b_0\Sigma_1 = e^{-\rho}B(x,\theta,\overline\theta)\overline c\overline\partial^2\overline c
\end{equation}
where $m$ is a four-dimensional index. $H^m$ should correspond to the supergravity superfield \cite{Siegel:1977ab,Siegel:1978mj}. The action \ref{hybrid massless action} reduces to
\begin{equation}\label{CY independent hybrid action}
    \frac{1}{8}\int d^4x d^4\theta [H^m D\overline D^2 D H_m + \overline D\overline\beta(2\partial_m H^m - \frac{1}{2} \overline D\overline\beta - D^2 B)]
\end{equation}
which has the gauge invariances
\begin{gather}
    \delta H^m = \overline D^2 \overline N^m + D^2 N^m + \partial^m P \\
    \delta \overline\beta = D^2\overline D P \\
    \delta B = 2\partial_m N^m - \frac{1}{2}\overline D^2 P + D\lambda \label{B gauge}
\end{gather}
and equations of motion
\begin{gather}
    D\overline D^2D H^m - \partial^m\overline D\overline\beta  = 0 \\
    \overline D_{\dot\alpha}(2\partial_m H^m - \overline D\overline\beta - D^2 B) = 0 \\
    D^2 \overline D\overline\beta = 0
\end{gather}
Although, as discussed above, the action is not real in Minkowski signature, the solutions to the equations of motion can be chosen to be. To see this, first note we can define $E = \overline D\overline\beta$ and work with the constrained field $E$, satisfying $\overline D^2 E=0$, instead of $\overline\beta$:
\begin{gather}
    D\overline D^2D H^m - \partial^m E = 0 \\
    \overline D_{\dot\alpha}(2\partial_m H^m - E - D^2 B) = 0 \label{eom 2}\\
    D^2 E = 0
\end{gather}
The third equation is the complex conjugate in Minkowski signature of the constraint $\overline D^2 E=0$, and the first equation above is manifestly real if $H^m$ and $E$ are. The second equation can be written as $2\partial_m H^m - E - D^2 B - \overline D^2\overline B = 0$ for some $\overline B$. So if we choose $B$ to be the complex conjugate of $\overline B$, this equation is also real. 
In the next section we will see how this gives a superspace description of four dimensional supergravity. After that, we will compare these results to the RNS formalism.

\subsection{Supergravity}
We expect the massless level of dimensionally reduced heterotic SFT to describe $N=1$ $d=4$ supergravity plus a tensor multiplet. One way to formulate this theory in four dimensional superspace is with the supergravity superfield $H_m$ plus a linear superfield \cite{Derendinger:1994gx}, i.e. a superfield satisfying $D^2E=0$ and $\overline D^2E=0$. The action is
\begin{equation}\label{sugra action}
    S = \frac{1}{8}\int d^4x d^4\theta[ H^m D\overline D^2D H_m + 2E\partial^m H_m - \frac{1}{2}E^2]
\end{equation}
which has the gauge invariances
\begin{gather}
    \delta H^m = \overline D^2 \overline N^m + D^2 N^m + \partial^m P \\
    \delta E = D\overline D^2D P
\end{gather}
and equations of motion
\begin{gather}
    D\overline D^2D H^m - \partial^m E = 0 \\
    D^2\overline D_{\dot\alpha}(2\partial_m H^m - E) = \overline D^2 D_{\alpha}(2\partial_m H^m - E) = 0 \label{sugra eom 2}
\end{gather}
This is equivalent to the equations obtained in the hybrid SFT since \ref{sugra eom 2} implies that $2\partial_m H^m - E = D^2 B + \overline D^2\overline B$ for some $B$ and $\overline B$. 

\subsection{Massless level in RNS}
Let us do a similar derivation using the RNS formulation of \cite{Berkovits:2004xh}, in order to compare. Here we work only in the NS sector. The string field has ghost number zero and picture zero. In the gauge $\xi_0\phi = 0$, the massless string field reads
\begin{gather}
    \Phi = c\xi e^{-\phi}\psi^M\overline c\overline\partial x^N A_{MN} + 2ic\partial c^+\xi\partial\xi e^{-2\phi}\overline c\overline\partial x^M \overline E_M + c\partial c^+\xi e^{-\phi}\psi^M E_M \nonumber\\
    +ic\xi\partial\xi e^{-2\phi}\overline c\overline\partial^2\overline c B + ic\xi\eta\overline B
\end{gather}
where $c^+ = \frac{1}{2}(c + \overline c)$ and the barred and unbarred fields are independent.

The linearized action is
\begin{gather}
    S = \int d^{10} x [A^{MN}(\frac{1}{4}\Box A_{MN} - \partial_M \overline E_N - \partial_N E_M) \nonumber\\
    - \overline E^M(\overline E_M + 2\partial_M\overline B) - E^M(E_M - 2\partial_M B) + 2\overline B\Box B]
\end{gather}
with gauge invariances
\begin{gather}\label{RNS massless action}
    \delta A_{MN} = \partial_N\xi_M + \partial_M\overline\xi_N \\
    \delta E_M = \frac{1}{2}\Box\xi_M + \partial_M f \\
    \delta \overline E_M = \frac{1}{2}\Box\overline\xi_M - \partial_M f \\
    \delta B = -\frac{1}{2}\partial^M\overline\xi_M + f \\
    \delta \overline B = \frac{1}{2}\partial^M\xi_M + f
\end{gather}

The RNS and hybrid formalisms can be related by field redefinitions \cite{Berkovits:1996bf}. This leads to the following identifications of the massless fields:
\begin{gather}\label{identifications}
    A_{mN} \propto D\sigma_m\overline D H_N |, \\
    A^j_N \propto \overline D^2\overline A^j_N |, \quad A_j^N \propto D^2 A_j^N |\\
    E_m \propto D\sigma_m\overline D \overline D\overline\beta |, \quad \overline E_M \propto D^2\overline D^2 H_M |, \\
    E_j \propto D^2\overline D^2C_j |, \quad E^j \propto D^2\overline D^2\overline C^j | \\
    B \propto D^2 B^{(hybrid)} |, \quad \overline B \propto \overline D\overline\beta | \label{identifications5}
\end{gather}
where the vertical bar denotes the $\theta=\overline\theta=0$ component. However, the hybrid formalism includes extra states with no RNS equivalent which will not be physical. To see this, first note that we can use $N^M$ and $\overline N^M$ in \ref{linear gauge 1} to gauge $H^M$ to Wess-Zumino gauge, where only the $\theta\overline\theta$ and $\theta^2\overline\theta^2$ terms survive for bosonic states. $\omega_{i\alpha}^M$ can be used to gauge away all of $A_i^M$ except the $\theta^2$ and $\theta^2\overline\theta^2$ terms. The latter has no RNS equivalent, but it can be easily checked that it is auxiliary by deriving the equation of motion from varying \ref{hybrid massless action} with respect to $\overline A_N^j$. The components of $\overline A_N^j$ work in an entirely analogous way. For $C_i$, we can use $\lambda_i$ and $\omega_{i\alpha}$ to gauge away everything except the $\theta^2\overline\theta^2$ term (this works similarly to a Wess-Zumino gauge, but we have an extra parameter to gauge away the $\theta\overline\theta$ term). $\overline C^i$ is analogous. $\overline\beta_{\dot\alpha}$ can be gauged, using $\alpha_{\dot\alpha}$, to a form where no terms with two $\overline\theta$'s appear. Among the remaining bosonic components, the $\theta\overline\theta$ term can be shown to be auxiliary by the $B$ equation of motion. Finally, $B$ can be gauged, by $\omega_{\alpha}$, to the form $\theta^2G(\overline y,\overline\theta)$, where $G$ is anti-chiral. The $\theta^2\overline\theta^2$ component can be shown to be auxiliary by the $\overline\beta$ equation of motion. All the remaining bosonic components of all fields are accounted for in \ref{identifications}-\ref{identifications5}. Thus, there are no extra physical states.

If we restrict to the CY independent states, it is easy to check that the RNS action
\begin{gather}
    S = \int d^{4} x [A^{mn}(\frac{1}{4}\Box A_{mn} - \partial_m \overline E_n - \partial_n E_m)\nonumber\\
     - \overline E^m(\overline E_m + 2\partial_m\overline B) - E^m(E_m - 2\partial_m B) + 2\overline B\Box B]
\end{gather}
and the hybrid action \ref{CY independent hybrid action} are compatible (in the NS sector) upon identifying the fields as in \ref{identifications}-\ref{identifications5}. Note that the hybrid field, even in the NS CY independent sector, still includes some extra states that do not correspond to any RNS state, but these are all pure gauge. Of  course, in addition, the hybrid field includes the R sector.

\section{Discussion}\label{sec Discussion}
In this paper we have constructed a linearized theory for heterotic string fields using the hybrid formalism. We found that the heterotic spectrum can be described in terms of three string fields, similarly to the open superstring. We can formulate the theory either in terms of ghost number $2$ fields annihilated by $\bar c_0$ or ghost number $1$ fields annihilated by $\bar b_0$. Reflecting these somewhat unusual constraints, the linearized equations of motion also take an unusual form involving the zero modes of $G^-$ and $\bar b$. We also analyzed explicitly the massless level. Although the theory is not real in Minkowski signature, we found that the equations of motion in the massless compactification independent sector can be written in a real form, and describe four-dimensional supergravity plus a tensor multiplet in superspace. If we include the CY states, since all the NS physical states coincide with the RNS formulation, and the action has manifest spacetime supersymmetry, the full linearized action must describe ten-dimensional supergravity in terms of four-dimensional superfields.

Having defined a linearized theory, we might expect to be able to construct an interacting theory following the ideas of \cite{Berkovits:2004xh} and \cite{Berkovits:1995ab}. However, given that the equations of motion now involve $G^-$ and $\bar b$ zero modes, the algebraic structures of the RNS formulation do not fully carry over. In particular, it is not clear how to construct string field products on which $G'^+_6$ and $\tilde G'^+_6$ would act in a simple way. We leave these questions for future work.

\section*{Acknowledgements}
NB would like to thank FAPESP grants 2021/14335-0, 2019/21281-4, 2019/24277-8 and CNPq grant 311434/2020-7 for partial financial support. UP thanks FAPESP grants 2022/13961-8 and 2023/00862-4 for financial support.

\bibliography{references.bib}

\providecommand{\href}[2]{#2}\begingroup\raggedright\begin{thebibliography}{10}

\bibitem{Berkovits:2004xh}
N.~Berkovits, Y.~Okawa and B.~Zwiebach, \emph{{WZW-like action for heterotic string field theory}}, \href{https://doi.org/10.1088/1126-6708/2004/11/038}{\emph{JHEP} {\bfseries 11} (2004) 038} [\href{https://arxiv.org/abs/hep-th/0409018}{{\ttfamily hep-th/0409018}}].

\bibitem{Okawa:2004ii}
Y.~Okawa and B.~Zwiebach, \emph{{Heterotic string field theory}}, \href{https://doi.org/10.1088/1126-6708/2004/07/042}{\emph{JHEP} {\bfseries 07} (2004) 042} [\href{https://arxiv.org/abs/hep-th/0406212}{{\ttfamily hep-th/0406212}}].

\bibitem{Berkovits:1995ab}
N.~Berkovits, \emph{{SuperPoincare invariant superstring field theory}}, \href{https://doi.org/10.1016/0550-3213(95)00259-U}{\emph{Nucl. Phys. B} {\bfseries 450} (1995) 90} [\href{https://arxiv.org/abs/hep-th/9503099}{{\ttfamily hep-th/9503099}}].

\bibitem{Zwiebach:1992ie}
B.~Zwiebach, \emph{{Closed string field theory: Quantum action and the B-V master equation}}, \href{https://doi.org/10.1016/0550-3213(93)90388-6}{\emph{Nucl. Phys. B} {\bfseries 390} (1993) 33} [\href{https://arxiv.org/abs/hep-th/9206084}{{\ttfamily hep-th/9206084}}].

\bibitem{Goto:2016ckh}
K.~Goto and H.~Kunitomo, \emph{{Construction of action for heterotic string field theory including the Ramond sector}}, \href{https://doi.org/10.1007/JHEP12(2016)157}{\emph{JHEP} {\bfseries 12} (2016) 157} [\href{https://arxiv.org/abs/1606.07194}{{\ttfamily 1606.07194}}].

\bibitem{Marcus:1983wb}
N.~Marcus, A.~Sagnotti and W.~Siegel, \emph{{Ten-dimensional Supersymmetric {Yang-Mills} Theory in Terms of Four-dimensional Superfields}}, \href{https://doi.org/10.1016/0550-3213(83)90318-8}{\emph{Nucl. Phys. B} {\bfseries 224} (1983) 159}.

\bibitem{Siegel:1977ab}
W.~Siegel, \emph{{Supergravity Superfields Without a Supermetric}},  HUTP-77/A068.

\bibitem{Siegel:1978mj}
W.~Siegel and S.J.~Gates, Jr., \emph{{Superfield Supergravity}}, \href{https://doi.org/10.1016/0550-3213(79)90416-4}{\emph{Nucl. Phys. B} {\bfseries 147} (1979) 77} HUTP-78/A019.

\bibitem{Derendinger:1994gx}
J.-P.~Derendinger, F.~Quevedo and M.~Quiros, \emph{{The Linear multiplet and quantum four-dimensional string effective actions}}, \href{https://doi.org/10.1016/0550-3213(94)90203-8}{\emph{Nucl. Phys. B} {\bfseries 428} (1994) 282} [\href{https://arxiv.org/abs/hep-th/9402007}{{\ttfamily hep-th/9402007}}].

\bibitem{Berkovits:1996bf}
N.~Berkovits, \emph{{A New description of the superstring}},  in \emph{{8th Jorge Andre Swieca Summer School: Particles and Fields}}, pp.~390--418, 4, 1996 [\href{https://arxiv.org/abs/hep-th/9604123}{{\ttfamily hep-th/9604123}}].

\end{thebibliography}\endgroup
\bibliographystyle{JHEP}

\end{document}